\documentclass[preprint,preprintnumbers,nofootinbib,aps,amsfonts]{revtex4}
\usepackage{verbatim}
\usepackage[hyperfootnotes=false]{hyperref}
\usepackage{graphicx}
\usepackage{epsfig}
\begin{document}
\title{Metastability in Anti de Sitter Space}
\author{Daniel Harlow}
\email {dharlow@stanford.edu}
\affiliation{Department of Physics, Stanford University, Stanford, CA, 94305}
\preprint{SU-ITP-10/11}
\begin{abstract}
I discuss conceptual issues associated with the presence of metastable AdS vacua in the string landscape.  The geometry of the decay from one AdS to another is presented in detail, and various subtleties that are not present for flat or dS vacua are demonstrated and analyzed.  I use mostly semiclassical gravity, but I consider the implications for recent attempts to construct field theory duals of metastable AdS vacua, as well as possible relevance to the study of eternal inflation. 
\end{abstract}
\maketitle

\section{Introduction}
Most discussions of the string landscape focus on either stable supersymmetric solutions, necessarily in flat or AdS space, or on unstable deSitter vacua.\cite{Bousso:2000xa,Kachru:2003aw}  It is widely expected that all flat vacua are supersymmetric and stable, but this is not the case for AdS vacua.  For example the ``uplifting'' of a supersymmetric AdS minimum can be incomplete, producing a SUSY-breaking vacuum with negative energy density.  These vacua are generically unstable, and we expect them to decay down to other AdS vacua.\footnote{In fact there are also more exotic decays which can take place, two examples of which I will discuss in section 5.  In the meantime I will focus on the more conventional Coleman-DeLuccia type of decay since most of the conclusions are identical for these more general decays.  For simplicity I will also sometimes work specifically in four-dimensions, but there is nothing intrinsically four dimensional about the issues I will discuss.}  Given our observation of a positive cosmological constant, these vacua are not of immediate phenomenological relevance.  Nonetheless they are still interesting for two reasons: firstly these vacua exist in the landscape, and they are part of the rich and poorly-understood dynamics of eternal inflation.  Any discussion of probability measures of eternal inflation will need to properly take into account the AdS vacua, and indeed recent measure proposals that do so have been found to predict that we should be living in AdS space!\cite{Salem:2009eh}  Clearly this prediction is wrong, but it demonstrates the danger of neglecting AdS vacua in a measure.  Since they must be included, we would do well to understand them.  Secondly, vacuum decay is an intrinsically non-perturbative phenomenon, either in string theory or in its low energy field theory, and understanding it completely requires a nonperturbative definition of the theory.  In flat backgrounds such definitions have only been found in supersymmetric spaces of high dimension, and in dS almost nothing is known.  By contrast AdS backgrounds are quite well understood via the AdS/CFT correspondence, and if the correspondence can be generalized to deal with metastable AdS vacua then it may suggest new approaches to thinking about quantum gravity in systems where the asymptotic regions of the theory fluctuate.  There has been some discussion of what a field theory dual of a metastable AdS might look like in the literature; I will discuss and elaborate on this work in later sections, but first I will present the precise thin-wall description of decays between AdS vacua.  
\section{Coleman-de Luccia Geometry in AdS space}
The effects of gravity on vacuum decay were first studied in the seminal paper of Coleman and de Luccia.\cite{Coleman:1980aw}  Using a saddle-point approximation to Euclidean gravity coupled to a scalar field, they presented a general method for calculating semiclassical decay rates of metastable vacua.  Their prescription can be easily summarized: look for SO(4) invariant Euclidean solutions of the four dimensional equations of motion which asymptotically approach the false vacuum, and then solve the Lorentzian equations of motion forward in time using a slice through the instanton as initial conditions.  The resulting geometry is interpreted as the growth of a critical droplet in a first-order phase transition, and the decay rate per unit volume is claimed to be:$$\Gamma\approx A e^{-S_B}$$ where $S_B$ is the action of the Euclidean instanton that produces the decay and $A$ is a dimensional prefactor.

For practical calculations it is very helpful to make use of Coleman's ``thin-wall'' approximation, valid when the height of the potential barrier is large compared to the difference in vacuum energies.  This is especially true when we are studying the geometry that evolves from the nucleation event: in this approximation we have a domain wall expanding outwards into the false vacuum, and the metric away from the wall is just that of dS, Minkowski, or AdS space depending on the value of the minima of the potential.  To the extent that this approximation is valid, the only knowledge the system retains of the scalar field dynamics is the values of the cosmological constant inside and outside the bubble and the tension of the domain wall.  We may then use the Israel junction conditions \cite{Israel:1966rt,Marolf:2005sr} to match the two geometries together and construct the full solution.  This procedure is explained especially clearly in \cite{Blau:1986cw}, so I will not review the derivation of the general formalism here.  

\begin{figure}[t!]
\centering
{\includegraphics[scale=0.6]{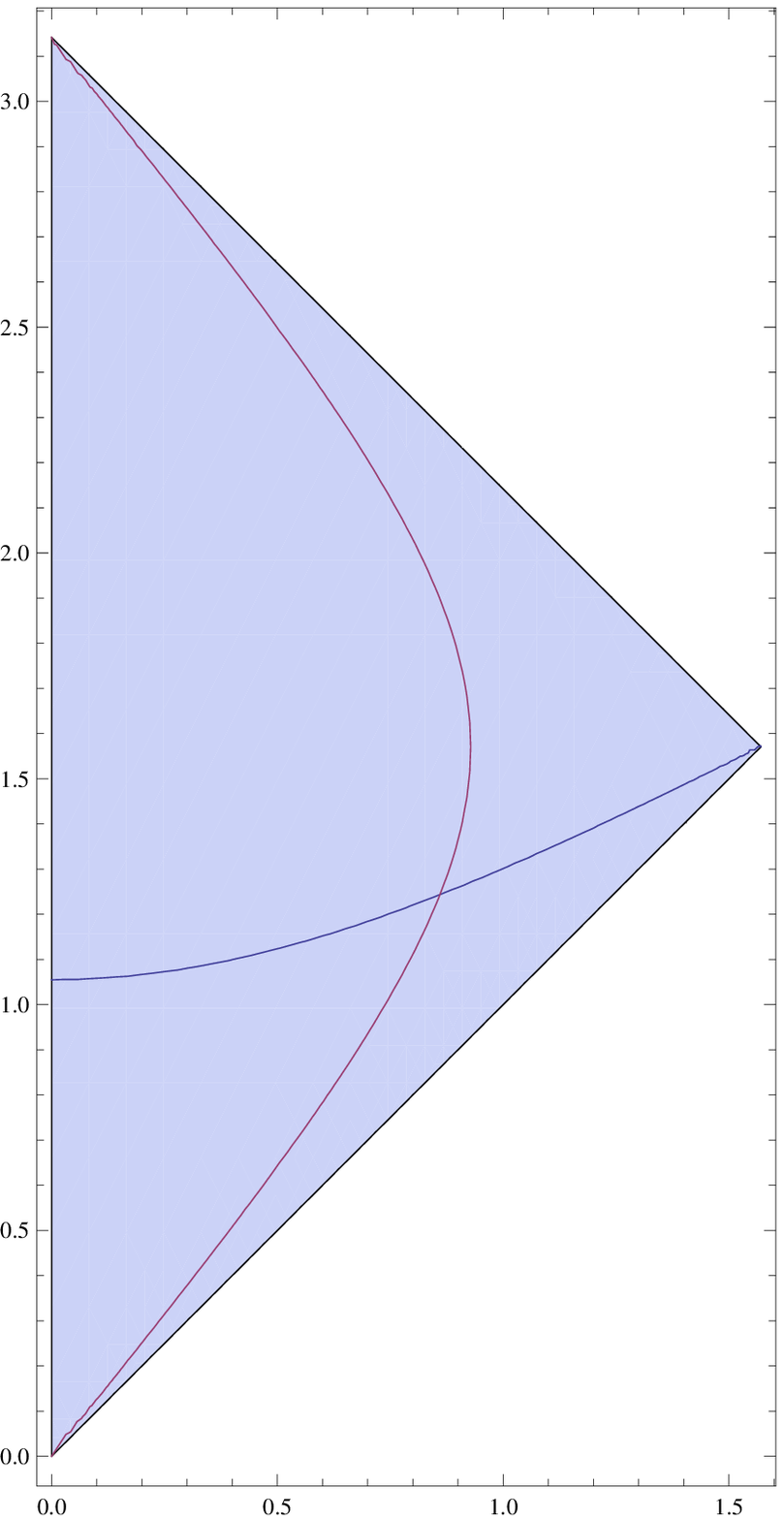}}\hspace*{10ex}
{\includegraphics[scale=0.612]{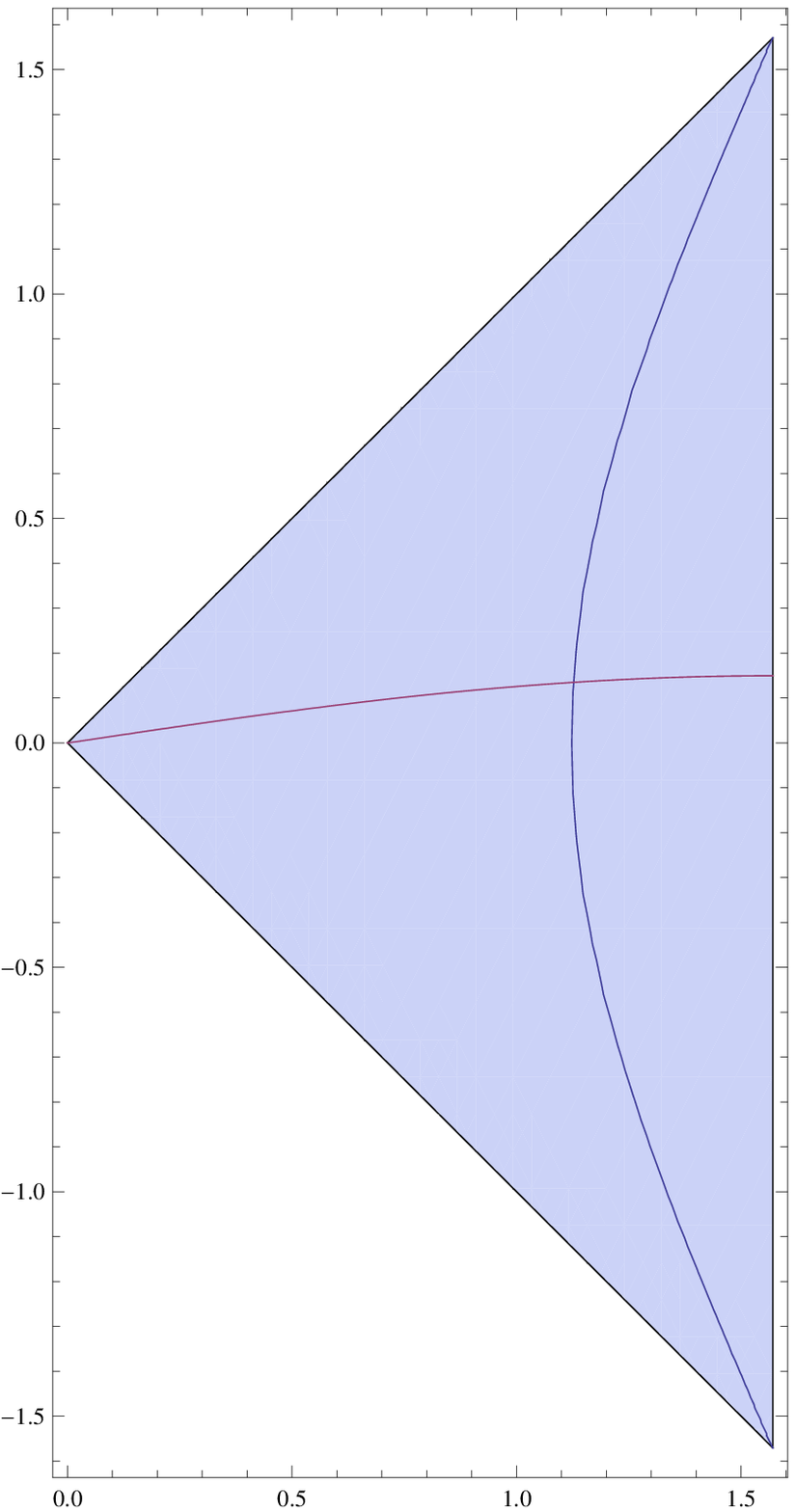}}
\caption{Location of inside and outside coordinate patches in the $\tau$-$\theta$ plane of global coordinates are shown on the left and right above.  $\theta$ is on the horizontal axis, and the boundary is at $\theta=\pi/2$ Lines of constant $\xi$ and $t$ are blue.}
\label{fig.1}
\end{figure}

To construct the Coleman de Luccia geometry in AdS space, it is very convenient to use a special set of hyperbolic coordinates that jointly cover one period of the entire space.  I will refer to these patches as ``interior'' and ``exterior''.  The interior coordinates are identical to an open FRW universe with a particular choice of scale factor:
\begin{equation} ds^2_{interior}=-dt^2+R^2 \sin^2(t/R) [dr^2+\sinh^2 (r) d\Omega_2^2]
\end{equation}
The exterior coordinates have a metric:
\begin{equation} ds^2_{exterior}=d\xi^2+R^2 \sinh^2(\xi/R)[-d\eta^2+\cosh^2(\eta) d\Omega_2^2]
\end{equation}
The advantage of these coordinates is that they make the SO(3,1) invariance of the Lorentzian solution manifest: the quantities in the square brackets are respectively the hyperbolic three plane for the interior coordinates and three dimensional de Sitter space in global slicing for the exterior coordinates, and each of these has isometry group SO(3,1).  So the AdS nature of the space enters only in the functions in front of the bracketed quantities.  I show the locations of these coordinates on the Penrose diagram of global AdS in figure \ref{fig.1}.\footnote{An analogous choice of coordinates also exists for flat and dS spaces, with the only difference being that one replaces $R\sin(t/R)\to R\sinh(t/R)$ and $R\sinh(\xi/R)\to R\sin(\xi/R)$ to get dS and replaces $R\sin(t/R)\to t$ and $R\sinh(\xi/R)\to \xi$ to get flat space.}
I will focus on the case of two AdS vacua, with radii $R_+$ and $R_-$ for the false and true vacua respectively.  Note that this means $R_+ > R_-$.  We are fortunate in that the high degree of symmetry makes the solution relatively easy compared to other situations that appear in the literature, for example in \cite{Freivogel:2007fx}.  In particular the trajectory of the domain wall must be invariant under SO(3,1), and since it is timelike it thus lies along a surface of constant $\xi$ in the exterior coordinate patch of each of the AdS vacua.  So we need only to determine the value of $\xi$ in each patch at which we make the joining, which we can do by solving the junction conditions.  For a general metric of the form: $$ds^2=d\xi^2+f(\xi)^2[-d\eta^2+\cosh^2(\eta) d\Omega_2^2]$$
and for which a domain wall of tension $\sigma$ lies along a surface of constant $\xi$, across which the energy density jumps by $\Delta \rho$, the four-dimensional junction conditions reduce to:

$$f_-(\xi_-)=f_+(\xi_+)\equiv f$$
\begin{equation}
f_-^{'}(\xi_-)-f_+^{'}(\xi_+)=4\pi G \sigma f
\end{equation}
$$f_-^{'}(\xi_-)+f_+^{'}(\xi_+)=\frac{2\Delta \rho f}{3 \sigma}$$

The first condition is the continuity of the metric across the domain wall, the second gives the appropriate discontinuity due to the energy density in the wall, and the third is energy-momentum conservation at the wall.  In fact the three are redundant, since energy momentum conservation follows from Einstein's equations, which give the first two conditions.  Now for the situation of interest we have $f_\pm(\xi_\pm)=R_\pm \sinh(\xi_\pm /R_\pm)$, so defining $x_\pm\equiv \xi_\pm/R_\pm$ we find: 

$$R_-\sinh(x_-)=R_+\sinh(x_+)$$
\begin{equation}
\cosh(x_-)-\cosh(x_+)=4\pi G \sigma R_-\sinh(x_-)
\end{equation}
$$\cosh(x_-)+\cosh(x_+)=\frac{2\Delta \rho}{3 \sigma}R_-\sinh(x_-)$$

We can check that the product of the last two equations, combined with the first, reduces to:
\begin{equation}
 \frac{1}{R_-^2}-\frac{1}{R_+^2} = \frac{8\pi G \Delta \rho}{3}
 \label{identity}
\end{equation}
which is indeed redundant.

To solve these conditions, we can look at the sum and difference of the second two equations, and defining $A\equiv 4\pi G \sigma R_-$ and $B\equiv \frac{2\Delta\rho}{3\sigma}R_-$, we find: $$\coth(x_-)=\frac{A+B}{2}$$ $$\coth(x_+)=\frac{R_+}{R_-}\frac{B-A}{2}$$
We can use equation (\ref{identity}) to write $B$ in terms of $A$, and then study when these conditions can be solved as a function of A.  The right hand side of the first equation is greater than one when $A>1+\frac{R_-}{R_+}$ or $A<1-\frac{R_-}{R_+}$, while the right hand side of the second equation is greater than one only when $A<1-\frac{R_-}{R_+}$.  Thus for there to be solutions for $x_\pm$ we must have $A \leq 1-\frac{R_-}{R_+}$, which gives an upper bound on the domain wall tension:\footnote{As far as I know this result first appears in \cite{Cvetic:1992st}, in the context of a BPS bound on domain walls in supergravity theories. It was also recently rederived in \cite{Narayan:2010em} by generalizing the CdL flat space derivation.  I am not aware of existing work that derives it from the junction conditions as I do here, but I expect someone has.}
\begin{equation}
\sigma\leq\frac{1}{4\pi G}\left(\frac{1}{R_-}-\frac{1}{R_+}\right)
\label{tensionbound}
\end{equation}
So we see that there is indeed a geometry that describes a bubble of AdS growing inside another AdS with larger radius, at least as long as the domain wall tension satisfies the bound (\ref{tensionbound}).  That this geometry does not exist when the tension violates the bound is the generalization to AdS space of the original insight of Coleman and de Luccia that it is possible for gravity to stabilize a purportedly metastable vacuum.  In particular we can observe that as $G\to 0$ with the energy densities fixed, we have the radii scaling like $\frac{1}{\sqrt{G}}$ and the bound becomes trivial.  
\section{Multi-Bubble Dynamics}
At the level of the geometry of the previous section, there is very little distinction in how the system works if we allow one or both of the vacua to be dS or flat instead of AdS.  The main difference is that when the false vacuum is dS, the junction conditions can always be solved and the decay is allowed for any value of the tension.  Also when the true vacuum is AdS it is widely expected that the coordinate singularity at $t=\pi R_-$ in the interior coordinates will be converted into a true singularity by SO(3,1) invariant fluctuations.  This is first claimed for the case of a flat false vacuum in the original Coleman-de Luccia paper, and their argument generalizes easily to the case where the false vacuum is AdS.  In particular corrections to the thin-wall approximation will generically involve rolling scalars inside the bubble, which will produce particles in an SO(3,1) invariant way that will then crunch the FRW universe.  I will have more to say about this later.

The physics becomes much more complex when we try to describe configurations with multiple bubbles.  Since we wish to follow Coleman in interpreting the presence of the instanton as giving a fixed decay rate per volume to produce such bubbles, we cannot avoid such situations.  For example let's consider the limit where we remove all gravitational effects, as we would when considering a metastable condensed matter system.  Bubbles will gradually form and expand, and for a while any particular bubble will be well-described in terms of an expanding domain wall with true vacuum inside and false outside.  But the bubbles will inevitably encounter one another, and they will eventually percolate and leave the system in a thermal state that looks nothing like the homogeneous true vacuum state.  It is only if we allow the system to dissipate heat into some ``environment'' that it can gradually relax to its ground state.  This makes it difficult to relate the decay rate to ``global'' observables, but we can accurately characterize the system by saying that the decay rate determines on average how long a particular observer will have to wait before being hit by a bubble.  We can be more precise about this: if we assume that at $t=0$ the system starts in a state with false vacuum everywhere, then the probability that a particular stationary observer will not be struck by any bubbles after time $t$ is:\footnote{To derive this equation, we demand that there are no nucleation events in the past light cone of the observer.  The probability that no bubbles nucleate on a particular spatial slice of volume $V$ through the past light cone is $\lim_{\epsilon \to 0}(1-dt\Gamma \epsilon^3)^{V/\epsilon^3}=e^{-\Gamma Vdt}$, and the total probability is a product of this over all times between $t=0$ and the current time.  Doing the calculation in a different dimension just changes the power of $t$ in an obvious way.  Also note that I have assumed the size of the critical bubble is negligible and we can treat domain wall trajectories as null.  In the appendix I discuss how to do such calculations in more detail.} 
\begin{equation}
P(t)=e^{-\Gamma \int_0^t{dt' \frac{4\pi t'^3}{3}}}\approx e^{-\Gamma t^4}
\label{probability}
\end{equation}
There are two interesting points to make about this formula:
\begin{itemize}
\item We can see that the probability is ``IR-safe'': it is independent of the volume of space.  This is as expected, since the velocity of the domain walls is limited by the speed of light.  For any finite time, there is a distance beyond which no bubble nucleated can reach a given observer, and we are free to truncate the system beyond that distance as we wish with no effect on the probability.  
\item By decreasing $\Gamma$, we can increase the average lifetime of a false vacuum observer without limit.  In fact I have not yet discussed the precise meaning of ``metastability'' but I would say that this property is as good a definition as any, at least semiclassically.  \end{itemize} 

If we now turn gravity back on, what happens depends strongly on the cosmological constants of the true and false vacuums.  The physics is most similar to the no-gravity case if the false vacuum is flat: then our formula (\ref{probability}) for the probability is still correct, as are our two observations.  Nonetheless the final fate of the system is quite different: rather than an innocuous thermalized state we expect gravitational collapse to a cosmological singularity!  This has not been proven rigorously, since the arguments cited above only solve Einstein's equations in the presence of a single bubble of AdS, but it seems unlikely that collisions of crunching bubbles could avoid crunching themselves.  At least for a single collision between two AdS bubbles this could probably be proven using the methods of \cite{Freivogel:2007fx}.  

Things become more interesting, both theoretically and phenomenologically, when the false vacuum is dS.  When the vacuum energy density is very small things proceed more or less as described above, with the false vacuum bubbles eventually percolating and quite possibly crunching.  But as we decrease the radius of the dS in Planck units while fixing the decay rate, we eventually reach a situation where the decay rate per Hubble volume and the Hubble expansion rate are comparable.\cite{Guth:1980zm}  The bubbles no longer completely percolate, and we move into the regime of eternal inflation.\cite{Linde:1986fc}  Nonetheless the two observations made above are still true, although the formula (\ref{probability}) no longer is.  More precisely, if we imagine fixing the dS radius and decreasing the decay rate, we can again make an average observer live as long as we like.  In fact it is easier than in flat space, since inflation pushes bubbles apart.  In this sense, dS is ``more metastable'' than flat space.  We can see this explicitly by noting that if we assume that at $t=0$ in the flat slicing of dS we have false vacuum everywhere, a computation similar to the one mentioned above (and done first I believe in \cite{Guth:1980zm})gives the late proper-time ($t\gg R_{+}$) probability for a stationary observer not to have collided with any bubbles as: $$P(t)\approx e^{-\Gamma t (R_{+})^3}$$
Thus we don't need to decrease $\Gamma$ as quickly with increasing $t$ to keep the probability large.  

This probability also remains volume independent, and in fact the situation in dS is again better than the situation in flat space.  Because our observer now sees a horizon, there are regions which can never nucleate bubbles that reach him/her/it, even in the limit of large time!  This statement is true in any spacelike slicing of dS.

We have found that at least with respect to the two above observations, metastable dS is better-behaved than metastable flat space.\footnote{This is contrary to the usual way the trouble flows as we change the cosmological constant.  There are several reasons for this, but one is that we are discussing a question that is independent of the future cutoff used to count observers within the nucleated bubbles, so many of the usual ambiguities of eternal inflation are not relevant here.  For measure enthusiasts however I will comment that these probabilities are calculated in an ensemble where observers are weighted by comoving volume, not physical volume.  The idea is to consider only observers who are already present at the initial surface, and then ask how long they survive.}  This suggests trouble when the false vacuum is AdS, and indeed in this case neither of the two above observations are true.  The crux of the issue is the well-known fact that massless particles in AdS can propagate out to the boundary and back in finite time.  One might hope that the presence of finite tension in the domain walls could save us from this conclusion, but we see from the explicit solution for the geometry that a domain wall of finite tension centered at the origin sweeps out a surface of constant $\xi$ in the external coordinates, which are asymptotically null and will indeed intersect any timelike geodesic in proper time which I will prove below is bounded above by $\frac{\pi R_+}{2}$.  Moreover since the volume of AdS is infinite, for a finite decay rate per unit volume such a bubble will always appear somewhere immediately.  This clearly violates the second ``metastability'' point, since any nonzero decay rate, no matter how small, will result in observers living at most for $\frac{\pi R_+}{2}$.\footnote{One might even be tempted to argue that this means that there is no such thing as a metastable AdS space, although this depends on how you define metastability.  I will continue to refer to AdS false vacuums which can decay by CdL instantons as metastable, but with this caveat in mind.  In particular we will see in section VI that these spaces are ``metastable enough'' to be considered valid landscape vacua.}  We can see that the ``volume-independence'' observation is also violated, since for $t\to \frac{\pi R_+}{2}$ an observer becomes sensitive to bubbles nucleated throughout entire spacelike slices of AdS.  So if we truncate the AdS space at some very large finite radius, as $t\to \frac{\pi R_+}{2}$ the probability to be struck by a bubble is suppressed more and more relative to the unregulated probability.  In the appendix I show that a timelike geodesic orthogonal to an initial spacelike slice of $AdS_{d+1}$ in global coordinates on which there are no bubbles has a survival probability: $$P(t)=\exp \left(-\frac{1}{d}\Gamma \cdot (R_+)^d V_{d-1} \int_0^t{d\tilde{t} \tan^d (\tilde{t} /R_+)}\right)$$ Here $V_{d-1}$ is the volume of a unit $S^{d-1}$, and t is proper time along the geodesic.  The integral can be done in specific dimensions or in general in terms of hypergeometric functions, but the point to notice is that it diverges as $t\to \frac{\pi R_+}{2}$.  So we see that indeed for any finite $\Gamma$ the probability to remain in the false vacuum goes to zero in a finite time bounded above by $\frac{\pi R_+}{2}$.\footnote{This probability formula is valid only for observers on geodesics that start out orthogonal to the initial surface, for whom the bound is saturated.  But I will show in the appendix that boosted geodesics also obey this bound and don't saturate it.}

So we are led to a rather unusual picture, in which observers in the false vacuum live for a finite proper time $t<\frac{\pi R_+}{2}$ and are then destroyed by a wave of bubbles coming in from all directions.  The aftermath is almost certainly singular, for the reasons recounted above, and decreasing $\Gamma$ only prolongs the inevitable momentarily.  

\section{A Field Theory Dual Description?}
The catastrophic dynamics described above constitute a serious qualitative change to the physics of gravity in a stable AdS background.  One might expect that the usual methods of AdS/CFT are not sufficient to describe them.  For one thing, in a metastable AdS space the positive energy theorem is not valid and there are states of arbitrarily negative energy in the Hilbert space.  So we expect that a proposed field theory dual would not have a stable ground state.\footnote{In infinite volume field theory there is a sense in which there can be a stable ground state while also having the Hamiltonian be unbounded below.  For example we could consider a theory where the effective potential has two non-degenerate minima.  The lower minimum is the stable ground state of the theory, but the energy difference between the two is IR divergent so if we define the energy of the higher metastable minimum to be zero, then the Hamiltonian appears to be unbounded from below.  In the situation we are interested in however, the field theory lives on a sphere of finite volume, so the Hamiltonian being arbitrarily negative means the effective potential itself would be unbounded from below.}  Additionally the expanding bubbles of true vacuum will reach the boundary of the space, destroying the asymptotic symmetries and in particular violating the conformal invariance usually present in the boundary theory.  Nonetheless one might hope that by appropriately regulating the theory, we may be able to describe an ``approximate'' dual field theory.  By cutting off the volume of the bulk AdS at some large radius, the volume for bubbles to nucleate in becomes finite, so by making the decay rate small in cutoff units we can extend the time it takes for a bubble to reach the regulated boundary to be quite long.  The energy would also become bounded from below.  In the proposed boundary theory this regulator will look like a UV cutoff on the field theory, and the pathologies will reappear as we try to take the continuum limit.  In this section I will explore this idea quantitatively and see what sense can be made of it.

First recall that in Poincare and global coordinates respectively, the metric of $AdS_{d+1}$ is:\footnote{In Poincare slicing, the units of $t$, $x$, and $z$ are ambiguous.  I will choose them to have units of length, but only their ratios are meaningful.} 

$$ds^2=\frac{R^2}{z^2}(-dt^2+dz^2+dx_{d-1}^2)$$
$$ds^2=\frac{R^2}{\cos^2\theta}(-d\tau^2+d\theta^2+\sin^2\theta d\Omega_{d-1}^2)$$
For stable AdS there is now very strong evidence that the bulk theory in Poincare slicing is dual to a conformal field theory on $\mathbb{R}^{d-1}$, while the bulk theory in global slicing is dual to the same conformal field theory on $S^{d-1}$.\footnote{I have listed here the spatial geometries, in both cases the \textit{spacetime} geometries have an additional factor of $\mathbb{R}$ for time.}\cite{Maldacena:1997re,Witten:1998qj}  We are interested in global bulk properties, so I will work almost exclusively in the global slicing.  

To do precise computations it is very convenient to introduce an IR cutoff at $\theta=\frac{\pi}{2}-\delta$, which is an $S^{d-1}$, and consider bulk dynamics only within it.  There is a well-known argument relating this IR cutoff to a UV cutoff in the boundary field theory~\cite{Susskind:1998dq}.  For readers unfamiliar with AdS/CFT, and to establish some conventions, the essence is as follows: in Poincare slicing one can use the AdS/CFT dictionary\cite{Gubser:1998bc,Witten:1998qj} to calculate field theory correlators at fixed separation in $t$ and $x$.  This calculation requires a regulator $z>\epsilon$, and the resulting correlation functions turn out to resemble CFT correlators only when the separations of operators in $x$ and $t$ are large compared to $\epsilon$.  There are additional $\epsilon$-suppressed corrections that become large when a separation approaches $\epsilon$, and which resemble the effects produced by adding irrelevant operators to the field theory Lagrangian.  There also turn out to be local divergences proportional to inverse powers of epsilon that can be removed by counterterms~\cite{Bianchi:2001kw,Skenderis:2002wp}.  So it is very natural to interpret $1/\epsilon$ as the UV cutoff of the field theory.  For global slicing, we can observe that for $\theta \to \frac{\pi}{2}$ the metric goes to $$ds^2=\frac{R^2}{\delta^2}(-d\tau^2+d\delta^2+d\Omega_{d-1}^2)$$
These coordinates are dimensionless, but we can redefine $\delta' \equiv \delta L$ and $\tau' \equiv \tau L$, where $L$ is some arbitrary length, to get:
$$ds^2=\frac{R^2}{\delta'^2}(-d\tau'^2+d\delta'^2+L^2 d\Omega_{d-1}^2)$$
Comparing this with the Poincare metric, we see that we can make the same argument as before, now interpreting $1/\delta'\equiv \Lambda$ as the UV cutoff.  The CFT lives on a sphere of radius $L$.\footnote{It is conventional but not required to choose $L=R$.}  The ratio of these two scales is $\delta$, which is independent of our arbitrary $L$, as it must be.  Thus we arrive at the global version of the famous ``UV/IR relation'': the radius of the bulk regulator sphere measured in units of the AdS radius is the same as the radius of the field theory sphere measured in cutoff units:
\begin{equation}
\frac{1}{\delta}=L\Lambda
\label{uvir}
\end{equation}
On the field theory side, this regulator will of course break conformal invariance, but it will be restored in the IR as we take the continuum limit.  The bulk dual of renormalization group flow is a procedure where we move the cutoff surface in towards the center, adjusting the boundary conditions to implicitly include the bulk physics in the spherical shell between the old and new cutoff surfaces, such that the low-energy boundary correlators (calculated at fixed-angle on the sphere) are preserved.  Using the dictionary we can see explicitly that changing the boundary conditions in the gravity theory does indeed correspond to deformation of the CFT action by integrals of local operators.\footnote{This description of renormalization group flow in the bulk is intrinsically vague, and the reason is that we currently do not have a real definition of gravity in AdS other than by the correspondence.  In particular what is meant by ``boundary conditions'' and ``boundary correlators'' in the bulk is not precisely defined, except in the large N limit where we can use classical gravity.  In that case, the ``integrating out'' procedure described here reduces to finding a solution of the classical equations of motion that satisfies the ``UV'' boundary conditions and then using its values on the lower energy cutoff surface as the ``IR'' boundary conditions.}

Using these results, we can see the problem with metastable AdS quite easily: if we imagine beginning evolution in global coordinates at $\tau = 0$ on a slice with no bubbles, any point on a regulator surface $\theta = \frac{\pi}{2}-\delta$ will be struck by bubbles from the boundary in global time $\tau \leq \delta$.  Since we have just learned that bulk effects from outside the regulator sphere are to be included as boundary conditions, this means that we must change the boundary conditions from the usual AdS/CFT ones within a field theory time $\tau' \leq \delta'$.  But we see from equation (\ref{uvir}) that this means the usual boundary field theory is only a good description for times less than its own UV cutoff!  A field theory that is only valid at times shorter than its own cutoff scale is a field theory that is of no use to anybody.  So it seems that any attempt to describe a metastable AdS space using a conventional boundary field theory is doomed from the outset.  One might object that bubbles reaching the boundary might be possible to describe in field theory terms, but we will see in the next section that this is rather unlikely.   

\begin{comment}
At this point I should comment that while I am assuming that bubbles striking the boundary invalidates use of the field theory, we need to be careful since even in stable AdS one encounters UV divergences that need to be removed by adding boundary counterterms to the bulk action.  One might hope that it is possible to absorb the effects of bubbles from the boundary into such counterterms, enabling a ``renormalized'' continuum description.
\end{comment}

Nonetheless in the literature there is an attempt to describe metastable AdS with a dual field theory which can live for a time that is parametrically longer than the lattice scale.\cite{Horowitz:2007pr}  I will discuss the details of this in the next section, but I will first discuss in general terms in what sense this is possible.  It is true that in order to give an accurate description of bulk physics in the vicinity of the cutoff surface we need to include the effects of bubbles from the boundary.  Even correlation functions that are widely separated on the field theory sphere will not accurately reflect the relevant bulk correlators near the cutoff surface.  But if we are willing to settle only for accuracy deep inside the bulk, we can imagine simply forbidding any bubbles from nucleating outside of the cutoff surface.  One ``physical'' way to do this is to simply cap off the AdS geometry with a compact manifold, for example a Calabi Yau.\cite{Kachru:2009kg}  This allows us to use the ordinary boundary conditions, with the restriction that we can then study only local observers far inside the regulator surface.  We learned in the previous section that such observers are not susceptible to signals from the boundary until their proper time is of order the AdS radius, so as we move the cutoff surface towards the boundary we can study observables for times approaching the AdS radius without being affected by the neglected dynamics outside the boundary.  But there is a crucial subtlety: as we move the cutoff towards infinity there is more and more volume for bubbles to nucleate in, and these bubbles will eventually reach the regulator surface from the inside.  Whether or not we can follow the boundary theory past this point is highly questionable, certainly our current understanding of the dictionary is much too primitive to handle bubble collisions with the boundary, so we had best restrict ourselves to times shorter than the time it takes for this to happen.  In the appendix I show that the probability that the regulator surface has not yet have been struck by any bubbles nucleated inside of it after global time $\tau$ is:
\begin{equation}
P(\tau) \approx \exp \left(-\frac{1}{d}\Gamma V_{d-1} R^{d+1} \tau \frac{1}{\delta^d} \right)
\end{equation}
From this formula we see, now adopting the convention $L=R$, that the critical field theory time is given by:
\begin{equation}
\tau_c' \approx \frac{1}{\Lambda (\Lambda R)^{d-1}(\Gamma R^{d+1})}
\label{irtime}
\end{equation}

Effective field theory computations should be accurate for times shorter than this but longer than the cutoff scale, and in particular we see that when $\Gamma R^{d+1}\ll \frac{1}{(\Lambda R)^{d-1}}$, which for finite cutoff is not hard to arrange since $\Gamma$ is usually exponentially surpressed, there is a large range of validity.  But we also see that there is an obstruction to increasing $\Lambda$ indefinitely while fixing $\Gamma$ and $R$, illustrating explicitly the problem with taking the continuum limit.  In fact for (\ref{irtime}) to give the correct late-time cutoff for the effective field theory, we also require that $\tau_c'< \pi R/2$, since as we approach this time even the center of the AdS is sensitive to the bubbles from outside the cutoff surface which we have ignored. This gives an additional inequality that bounds the decay rate the other way:
$$\Gamma R^{d+1}>\frac{1}{(\Lambda R)^d}$$
The first inequality guarantees that the field theory will be valid for times large compared to its cutoff, while the second ensures that equation (\ref{irtime}) gives the correct bound.  When the second inequality is violated we need to instead use $R \pi/2$ as the late-time cutoff.  Note however that for $(\Lambda R)\gg 1$, which is required for the field theory on the sphere to make sense, there is a wide range for $\Gamma R^{d+1}$ to satisfy both inequalities.  Regardless we conclude that for times less than whichever of these two late-time cutoffs is shorter, we should be able to study local bulk physics using an effective field theory dual.  Of course as of now tools for answering local bulk questions using the field theory dual are rather limited, and indeed the only things that we really know how to calculate are exactly the boundary correlators which will not be accurately computed using these methods.  To make this proposal practical for computations would require a major advance in our understanding of the bulk/boundary duality even in the usual stable AdS setup.  Some proposals in this direction are made in \cite{Susskind:1998vk,Polchinski:1999ry,Polchinski:1999yd}, and more recently by \cite{Gary:2009mi} but much remains to be understood.  

In closing this section, I will observe that there is an important difference between the two proposed late-time cutoffs.  The one given by equation (\ref{irtime}) arises from bubbles nucleating inside of the cutoff surface and propagating outwards, and we should be able to see this pathology develop in the boundary field theory at late times.  We will see some evidence of this in the next section.  By contrast the late-time cutoff at $\pi R/2$ should be invisible in any proposed field theory dual, since in using the field theory at all we have neglected the physics which gives rise to it.  

\section{Discussion of Existing Work}
It is now long past time to clarify which of the above ideas are already present in the literature, and to see to what extent my discussion clarifies or benefits from existing understanding.  The key papers in which some of these issues are discussed are \cite{Hertog:2004rz,Hertog:2005hu} and especially \cite{Horowitz:2007pr}.  In \cite{Horowitz:2007pr} the authors consider the AdS/CFT correspondence on $AdS_5 \times S^5/Z_k$, showing that the bulk gravity theory can undergo a ``bubble of nothing'' decay as first described in \cite{Witten:1981gj}, and moreover that from the perspective of the false vacuum AdS the transition (for large $k$) is basically identical to a more typical thin-wall CdL process.  The authors identify that the infinite volume of global slices of the AdS means that the decay rate in the field theory will be UV divergent.  They then claim that it should be possible to introduce an effective field theory dual that is valid for times longer than the UV cutoff and shorter than an IR cutoff given (using my notation) by:
\begin{equation} 
\tau_c'\approx \frac{1}{\Lambda (\Lambda R)^3 (\Gamma R^{5})}
\label{hop}
\end{equation}
Here $\Gamma$ is the five dimensional decay rate, related to the ten-dimensional decay rate by $\Gamma=\Gamma_{10} R^{5} V_5 /k$, where $V_5$ is the volume of a unit $S^5$ and the factor of $k^{-1}$ incorporates the orbifolding.  In \cite{Horowitz:2007pr} it is shown that the ten dimensional decay rate is given by $\Gamma_{10}\approx e^{-B} (\frac{k}{R})^{10}$, where $B$ is the action of the relevant ten dimensional instanton, and substituting these into equation (\ref{hop}) the reader can reproduce the result for $\tau_c'$ given in their paper:\footnote{This is for what the authors refer to as the ``localized'' bounce, which occurs at a definite point in $S^5/Z_k$.  There is also a ``smeared'' bounce, but its decay rate is smaller.}
$$\tau_c'=\frac{e^B}{\Lambda (k^3 \Lambda R)^3}$$
The authors of \cite{Horowitz:2007pr} arrived at this long-time cutoff by estimating how long in global time it takes for a single bubble to nucleate anywhere inside the cutoff surface.  This is not quite the same calculation as I did in the previous section, but nontheless the cutoff (\ref{hop}) is actually equal to the one from (\ref{irtime}) that I proposed above.  More accurately as shown in the appendix it is equal up to subleading contributions in a large $1/\delta$ expansion.  My analysis differs from (and I believe improves on) theirs in three respects.  First in the realization that this effective field theory can only be used to study physics deep in the bulk.  Boundary correlators, even those calculated in the allowed region of global time and at wide angular separation, do not accurately describe the bulk physics near the cutoff surface.  In field theory terms, they are sensitive to the ``capping off'' of the geometry in the UV, and do not take into account the ``integrated out'' effects of bubbles nucleated outside of the cutoff surface.  Secondly I have motivated their late-time cutoff in a different and I believe more physical way: in deriving (\ref{hop}), the authors assumed that the field theory description is invalid as soon as a bubble nucleates anywhere inside, whereas I considered it to be good until a bubble actually hits the cutoff surface.  That these different analyses produced the same answer is a consequence of the fact that most of the volume in AdS is near the boundary, and the first bubble that nucleates is most likely in the vicinity of the boundary and will hit it soon after.  The field theory should be able to describe the growth of bubbles until they hit the boundary, and it is only the fact that nucleation and arrival at the boundary happen almost simultaneously that the two calculations agree.  Finally I claim that the late time cutoff (\ref{hop}) is only valid when it is less than $\pi R/2$, but that as discussed above this effect should be invisible in the field theory dual.  

As evidence that the field theory should indeed be able to describe the growth of bubbles, we can turn to \cite{Hertog:2004rz,Hertog:2005hu}.  These authors study scalar fields on $AdS_4$ and $AdS_5$ but with generalized boundary conditions that correspond through the dictionary to adding multitrace operators to the field theory action.  These generalized boundary conditions allow for instanton solutions which continue to expanding thick-wall bubbles inside of which the scalar field is rolling down to negative infinity.  The authors observe that a single such bubble takes global time $\tau=\pi/2$ to reach the boundary, and propose that in the dual field theory this process is described by a homogeneous (boundary) scalar field expectation value rolling down an unbounded effective potential and reaching infinity at finite time $\tau'=R \pi/2$.  They interpret this as being dual to the formation of a gravitational singularity in the bulk.  To the extent that these models are correct, they suggest that it is indeed possible to describe the formation and expansion of bubbles in the boundary effective field theory, at least until the bubbles reach the boundary.  However this present work raises a potential problem with this literature.  The claim that the boundary scalar field rolls down to negative infinity in field theory time $\tau'=R \pi/2$ seems to contradict the above claim that the field theory breaks down as the time approaches the upper bound (\ref{irtime}), which in the continuum regime they are interested in will be much less than $\pi R/2$.  The issue is that the authors of \cite{Hertog:2004rz,Hertog:2005hu,Craps:2007ch,Craps:2009qc} consider only single-bubble solutions and neglect the possibility of further nucleation of more bubbles.  It is easy to check that their boundary conditions admit both zero and one-instanton solutions (zero-instanton means empty AdS), and the one-instanton solutions can be moved around using the asymptotic symmetries without violating the boundary conditions.  So we might expect that we can superpose two widely separated such instantons and slice through their common center to find a find a legitimate multi-bubble configuration.  If their theory allows the nucleation of multiple bubbles, then their picture of a scalar field smoothly rolling down to negative infinity will be wrong, since already after time $\tau_c'$ their regulator surface will start seeing bubble collisions.  In fact I believe that in many cases their boundary conditions do forbid multi-instanton solutions.  Consider for example the $AdS_4$ model discussed in \cite{Hertog:2004rz,Hertog:2005hu}.  In coordinates where the $AdS_4$ metric is: 
$$ds^2=-(1+r^2)dt^2+\frac{dr^2}{1+r^2}+r^2 d\Omega_2^2$$
the authors impose boundary conditions on the scalar field:
$$\phi(r,t,\theta_i)\to \frac{\alpha(t,\theta_i)}{r}+\frac{f \alpha^2(t,\theta_i)}{r^2}+\mathcal{O}(\frac{1}{r^3})$$
Here $f$ is a fixed constant that is the same for any field configuration while $\alpha(t,\theta_i)$ is an arbitrary function of the nonradial coordinates.  It is easy to see that if we superimpose two solutions that obey these boundary conditions with different $\alpha$'s, the superposition obeys the boundary conditions only if the alphas have no common support.  Since the $\alpha$'s for their instantons are everywhere nonvanishing, this suggests that there are no multi-instanton solutions that respect the boundary conditions, although it does not prove it.  If we assume such a proof exists, then this model gives an explicit example where the field theory description is valid even after the single allowed bubble nucleates, thus providing support for the philosophy and (slightly) later late-time cutoff I advocate above.  In fact in this case the lack of other instantons suggests that we can effectively take the continuum limit without encountering any problems until the bubble actually reaches the boundary at $\tau=\pi/2$.\footnote{An interesting question: if these theories indeed do not allow multi-instanton solutions, what does this mean for the usual CdL calculation of the decay rate?}  

Before closing, I will briefly comment that this superposition argument seems to fail for the boundary conditions used in the $AdS_5$ model of \cite{Craps:2007ch}:
$$\phi(r,t,\theta_i)\to \frac{f \alpha(t,\theta_i)\log(r)}{r^2}+\frac{\alpha(t,\theta_i)}{r^2}+\mathcal{O}(\frac{1}{r^3})$$
Here again $f$ is a specified constant and $\alpha$ is arbitrary, but we can see in this case that superpositions of widely spaced instanton solutions will still obey the scalar boundary conditions, and it seems that multi-instanton solutions exist and invalidate the picture advocated in \cite{Craps:2007ch}.  However their boundary conditions for the metric are actually quadratic in $\alpha$, so the combined metric/scalar boundary conditions probably stabilize their setup against multi-bubble configurations.
\section{Relation to Eternal Inflation}
So far my discussion has been quite formal, and the reader may be wondering what relevance it has for real-world physics.  In recent years theoretical and some indirect experimental evidence has accumulated that our observable universe is a small part of a vast and nonhomogeneous megaverse populated by the mysterious mechanism of eternal inflation~\cite{Susskind:2003kw}.  The string landscape \cite{Bousso:2000xa,Kachru:2003aw} appears to contain many AdS vacua, and most of them can probably be nucleated from a parent de Sitter.  Some of these AdS vacua will be metastable, for a recent discussion see \cite{Narayan:2010em}, and since they are populated by eternal inflation we will need to deal with them if we wish to make statistical predictions.  As pointed out in the introduction, there are otherwise well-regarded measures which actually predict we live in AdS space~\cite{Salem:2009eh}.  The fact that many of these AdS vacua are unstable and kill all their observers within an AdS time seems like it would have a dramatic effect on such predictions, but in fact it does not.  The reason is simply that even bubbles of stable AdS vacua nucleated from eternal inflation are already unstable to the crunch described at the beginning of section III.  So the measures under consideration, for example in \cite{DeSimone:2008bq}, already assume that their AdS observers\footnote{Referred to by Freivogel as our ``evil twins''} only live for an AdS time!  So in this regard the issues discussed in this paper are irrelevant to what could (somewhat controversially) be viewed as actual physics.  

More formally however, we have illustrated something which I believe to be important.  Efforts to make probabilistic predictions in an eternally inflating megaverse have been fraught with ambiguity, and attempts to formulate holographic descriptions of dS space \cite{Strominger:2001pn,Witten:2001kn} have met with minimal success.  These problems are argued by Susskind to be related, and are claimed by him to arise from what he calls the ``asymptotically warm'' nature of de Sitter space.\footnote{Actually he would also include the presence of horizons as a source of additional trouble, but unfortunately the present work has nothing illuminating to say about this problem.}  This means that bulk fields, including gravity, fluctuate all the way out to spatial infinity, and there is no clear boundary where one might define a non-gravitational holographic theory.  This is to be contrasted with stable AdS space or eleven-dimensional flat space, where such a description is possible.  However we have seen here that a metastable AdS space does indeed fluctuate all the way out to its boundary, but yet it admits an approximate holographic description motivated by the AdS/CFT correspondence.  Observations deep inside the bulk can be studied using a non-gravitational effective field theory that is valid over timescales that can be quite long.  Indeed this possibility was the original motivation for the present work.  However the situation is less nice than it appears: we found above that the field theory is only valid until a time that is basically the same as the time when the first bubble forms somewhere in the bulk.  Given that this bubble is most likely to form near the cutoff surface, an observer sitting in the center will be very unlikely to see it.  So even though the field theory dual can in principle describe the nucleation and growth of multiple bubbles, the geometry of AdS space conspires to ensure that it breaks down well before a typical observer starts to see them!  The reader can confirm this by evaluating formula (\ref{observer}) at time $\tau_c'$.  So the verdict is once again that asymptotic warmness seems to prevent a conventional holographic description in the region of interest for eternal inflation.

\section{Conclusion}
I have now covered a lot of ground, and it may be useful for the reader to see a summary of the points which seem to me to be original.
\begin{itemize}
\item[(1)]  The property of metastability, in terms of observations made by local observers, is actually best defined in dS space.  In flat space it still ``barely'' satisfies the two desireable properties defined in section III, but in AdS space neither is satisfied and for any nonzero decay rate there is a firm limit to how long any observer can live.
\item[(2)] Any attempt to define an effective boundary field theory of a metastable AdS will only succeed when computing observables located deep in the bulk.  In particular the boundary correlators calculated using the usual dictionary will not accurately describe physics in the neighborhood of the cutoff surface, which in fact will be obliterated by bubbles from the boundary within a single lattice time.  
\item[(3)] The lifetime of validity for this effective field theory dual proposed in \cite{Horowitz:2007pr} is only correct when it is less than $\pi R/2$, and even then it should be slightly extended to include an approximate description of bubble nucleation and growth.
\item[(4)] The fact that the effective field theory breaks down at time $\tau_c'$ should be visible in the boundary theory, but in the region where this time is longer than $\pi R/2$ the fact that the bound is no longer valid should not be visible since we ignored it by definition by assuming a valid field theory description.
\item[(5)] The work of \cite{Hertog:2004rz,Hertog:2005hu,Craps:2007ch,Craps:2009qc} succeeds only to the extent that their boundary conditions prevent multi-bubble configurations from forming.
\item[(6)] Metastable AdS vacua are perfectly valid members of the landscape, and indeed measure proposals can treat them as effectively stable for the purposes of making statistical predictions (provided that they are assumed to crunch).
\item[(7)] The effective field theory description first advocated in \cite{Horowitz:2007pr} and expanded on here, provided it could really be constructed, would be the first example (as far as I know) of a holographic description of a background that fluctuates asymptotically.  Unfortunately it succeeds in this only because it becomes invalid before it becomes interesting.  
\end{itemize}
There are still many open questions, in particular:
\begin{itemize}
\item[(1)] Is there an explicit description of the effective field theory dual?  For the instability studied in \cite{Horowitz:2007pr}, the answer should be that it is just a target space orbifold of the usual $\mathcal{N}=4$ super Yang-Mills theory by a discrete freely-acting subgroup of its $SO(6)$ R-symmetry, and in particular the negative energy states must lie in the twisted sector of the field theory, since the ordinary $\mathcal{N}=4$ states obviously have positive energy.  It would be interesting to explore this in more detail, and for example to try to reproduce the late-time cutoff.  Of course the orbifolding breaks all the supersymmetry, so at large $\lambda$ any precise study would be difficult.  It would also be interesting to find an effective field theory description of a pure CdL AdS-AdS decay, and in particular to see if the lower AdS appears as some sort of resonance.\footnote{Several days before this paper was submitted, \cite{Barbon:2010gn} appeared which claims to propose such a dual.  It will be interesting to see to what extent these issues can be addressed in this theory.} 
\item[(2)] Is there an unstable brane system whose near-horizon physics resembles a metastable AdS space?  If so then we could study it using bulk string theory as a regulator and understand to what extent it can be described by an effective field theory.  It would be interesting to reproduce the late-time cutoff in this language, and also to understand to what extent gravity does or does not decouple at times later than it.  What I expect will happen is that the fact that bubbles reach the boundary in finite time really reflects the fact that in an unstable brane configuration the near-horizon region never really decouples from the asymptotically-flat ten dimensional bulk, and bubbles that reach the boundary ``leak out'' into the asymptotically flat region.  This would mean that any attempt to describe the near-horizon region as being a closed Hamiltonian system would break down as the system decays, and indeed this is what is found.
\item[(3)]  Can we learn anything relevant for eternal inflation from these ideas?  The prognosis is not good for the reasons discussed at the end of section VI, but it may be that the mere fact that bulk cosmological observations in the presence of a bubbling asymptotically warm geometry can be calculated in a non-gravitational effective field theory is already useful.  In particular the failure of this proposal to describe observers who see bubbles was really due to particular facts about the geometry of AdS space.  As I discussed above, metastability is the ``best'' in dS space, and perhaps we can use this.  
\item[(4)] What is the real quantum gravity description of a metastable AdS that allows accurate description of physics all the way out to the boundary?  The fact that bubbles from outside the regulator surface reach it almost immediately suggests that it is possible that gravity does not decouple.  This has been argued to be the case in other examples of asymptotically warm backgrounds \cite{Freivogel:2006xu,Polchinski:2009ch}, and the brane picture just given seems to confirm this.  However even the brane picture is still really a type of regulator, with the full nonperturbative description being in 10-dimensional flat space.  Perhaps there is no satisfactory definition of physics exclusively in metastable AdS space, and that it appears only as an approximation to an unstable brane or as a crunching byproduct of eternal inflation.  
\end{itemize}
\acknowledgments
I'd like to thank Alex Westphal, Vitaly Vanchurin, Xi Dong, Sandip Trivedi, Eliezer Rabinovici, and especially  Mahdiyar Noorbala and Stephen Shenker for useful discussions.  I would also like to thank my adviser Leonard Susskind both for extremely helpful conversations about many aspects of this work, and also for encouragement to pursue my own interests and patience while I do so.  During this work I have been supported both by a Melvin and Joan Lane Stanford Graduate Fellowship and also by the Stanford Institute for Theoretical Physics and the Stanford Physics Department.

\appendix
\section{Nucleation Probabilities}

In this appendix I will derive the various probability formulas quoted in the text.  The basic idea is that if we assume that there is some initial spacelike slice through a spacetime on which there are no bubbles and the geometry is purely that of the false vacuum, then at later times we can find the probability that some point (or set of points) is still in the false vacuum by calculating the probability that no bubbles have nucleated anywhere in the part of its past light cone that lies above the initial surface.  If we discretize this region, which I'll denote $\mathcal{C}$, into smaller regions of infinitessimal spacetime volume $dV_i=\sqrt{-g_i}d^dx$, where $i$ runs over the discrete regions, the probability that no bubbles nucleate is:
$$P(\mathcal{C})=\prod_{i\in \mathcal{C}} (1-\Gamma dV_i)$$
We now take the continuum limit by increasing the number of discrete regions $N$ and decreasing their size $dV_i$ such that the volume $NdV_i$ stays finite.  To disentangle the various cross terms we can take the log:
$$\log{P(\mathcal{C})}=\sum_i \log(1-\Gamma dV_i)=-\Gamma \sum_i (dV_i+\mathcal{O}((dV_i)^2))$$
So in the $N\to \infty$ limit we find:
\begin{equation}
P(\mathcal{C})=e^{-\Gamma \int_{\mathcal{C}} dV}=e^{-\Gamma \int_{\mathcal{C}} \sqrt{-g}d^d x}
\end{equation}

Let's first use this to derive the probability that an observer on a timelike geodesic orthogonal to the initial surface has not yet been struck.  In global coordinates the metric of $AdS_{d+1}$ is:
$$ds^2=\frac{R^2}{\cos^2{\theta}} \left(-d\tau^2+d\theta^2+\sin^2{\theta}d\Omega_{d-1}^2\right)$$
and we have:
$$\sqrt{-g}=R^{d+1}\frac{\sin^{d-1}{\theta}}{\cos^{d+1}{\theta}}$$
I will of course choose coordinates so that the initial slice lies at $\tau=0$, and I will use the symmetries to move the geodesic in question to the center.  Since it is orthogonal to the initial surface, it will just sit at the origin.  Radial null geodesics in these coordinates are just diagonals in the $\tau-\theta$ plane, so the lightcone is simple to describe and we find:
\begin{eqnarray*}
P(\tau)&=&\exp \left(-\Gamma \int_0^\tau d\tilde{\tau} \int_0^{\tilde{\tau}}d\theta \int d\Omega \sqrt{-g}  \right)\\
&=&\exp \left(-\Gamma R^{d+1} V_{d-1} \int_0^\tau d\tilde{\tau} \int_0^{\tilde{\tau}}d\theta \frac{\sin^{d-1}{\theta}}{\cos^{d+1}{\theta}} \right)\\
&=&\exp \left(-\frac{1}{d}\Gamma R^{d+1} V_{d-1} \int_0^\tau d\tilde{\tau} \tan^d{\tilde{\tau}} \right)
\end{eqnarray*}
So rewriting in terms of the proper time along the geodesic $t\equiv \tau R$, we get:
\begin{equation}
P(t)=\exp \left(-\frac{1}{d}\Gamma R^{d} V_{d-1} \int_0^t d\tilde{t} \tan^d(\tilde{t}/R) \right)
\label{observer}
\end{equation}
which is the equation quoted in the text.  For geodesics which are boosted with respect to the initial surface a computation of this sort would be more difficult, but fortunately there is no need.  Timelike geodesics starting in the center of AdS space with nonzero velocity fly out towards the boundary but eventually fall back in a proper time which turns out to be independent of the initial velocity and is equal to $\pi R$.  They also return after the same amount of global time, which is always $\tau=\pi$.  By symmetry this means that the turning point, which happens at $\tau=\pi/2$ in global coordinates, happens at proper time $\pi R/2$.  But at this point the geodesic would have already passed through the shell of bubbles coming in to obliterate the center of the space at global time $\tau=\pi/2$, and thus would have already decayed.  So the upper bound on the lifetime of timelike observers cannot be circumvented by boosting them, as promised above in the text.

We can similarly derive the probability for the boundary at $\theta=\pi/2-\delta$ not to have been struck by any bubbles after global time $\tau$, the relevant integral is:
\begin{eqnarray*}
P(\tau)&=&\exp \left(-\Gamma V_{d-1} R^{d+1} \int_0^\tau d\tilde{\tau} \int_{\pi/2-\delta-\tau+\tilde{\tau}}^{\pi/2-\delta} d\theta \frac{\sin^{d-1}{\theta}}{\cos^{d+1}{\theta}} \right)\\
&=&\exp \left(-\frac{1}{d}\Gamma V_{d-1} R^{d+1} \left[\tau \tan^d(\pi/2-\delta)- \int_0^\tau d\tilde{\tau}\tan^d(\pi/2-\delta-\tau+\tilde{\tau})\right]\right)\\
&=&\exp \left(-\frac{1}{d}\Gamma V_{d-1} R^{d+1} \left[\tau \frac{1}{\delta^d} + \mathcal{O}(\frac{1}{\delta^{d-1}})\right]\right)
\end{eqnarray*}
In the last line I have assumed $\delta\ll \tau \ll \pi/2$, consistent with the discussion in the text, and used some facts about hypergeometric functions.  We thus see that in this regime we have:
\begin{equation}
P(\tau) \approx \exp \left(-\frac{1}{d}\Gamma V_{d-1} R^{d+1} \tau \frac{1}{\delta^d} \right)
\label{boundaryhit}
\end{equation}
This is the result used above in the text.

Finally in \cite{Horowitz:2007pr} the authors estimate the expected time before a bubble nucleates anywhere within the cutoff surface.  To reproduce this result, we can calculate the probability that no bubbles have nucleated between the the initial surface and a surface of constant global time $\tau$.  The integral is the same as the previous one, but with the $\theta$ integral now extending down to zero.  This only affects the subleading terms in $\frac{1}{\delta}$, so the expression for the probability is still given at leading order by equation (\ref{boundaryhit}).
\bibliography{adsdecaypaper}

\end{document}